# DAS-MAE: A self-supervised pre-training framework for universal and high-performance representation learning of distributed fiber-optic acoustic sensing

Junyi Duan, Jiageng Chen, and Zuyuan He *

State Key Laboratory of Photonics and Communications, Shanghai Jiao Tong University, Shanghai 200240, China.

**Abstract:** Distributed fiber-optic acoustic sensing (DAS) has emerged as a transformative approach for distributed vibration measurement with high spatial resolution and long measurement range while maintaining cost-efficiency. However, the two-dimensional spatial-temporal DAS signals present analytical challenges. The abstract signal morphology lacking intuitive physical correspondence complicates human interpretation, and its unique spatial-temporal coupling renders conventional image processing methods suboptimal. This study investigates spatial-temporal characteristics and proposes a self-supervised pre-training framework that learns signals' representations through a mask-reconstruction task. This framework is named the DAS Masked AutoEncoder (DAS-MAE). The DAS-MAE learns high-level representations (e.g., event class) without using labels. It achieves up to 1% error and 64.5% relative improvement (RI) over the semi-supervised baseline in few-shot classification tasks. In a practical external damage prevention application, DAS-MAE attains a 5.0% recognition error, marking a 75.7% RI over supervised training from scratch. These results demonstrate the high-performance and universal representations learned by the DAS-MAE framework, highlighting its potential as a foundation model for analyzing massive unlabeled DAS signals.

## Introduction

Distributed fiber-optic sensing (DFOS) techniques have undergone rapid advancement and widespread applications in recent years [1-3], especially distributed acoustic sensing (DAS) emerged as a preeminent way for strain measurement [4-5]. Based on phase-sensitive optical time-domain reflectometry ($\Phi-\mathrm{OTDR}$) [6], a DAS interrogator analyzes the phase of Rayleigh backscattering light along the sensing fiber. Thus, the dynamic strain information can be acquired. It enables distributed measurements by converting the sensing fiber that spans up to hundred kilometers into an array of vibration sensing nodes (similar to microphones) located as close as a few meters apart. Compared to conventional electromechanical sensors, DAS yields distinct advantages including high resolution, small size, power-free operation, and adaptability to harsh environments. These advantages have driven widespread deployment across earthquake monitoring [7-8], oil and gas pipeline monitoring [9-10], and railway monitoring [11-12]. However, DAS-acquired spatial-temporal data (often referred to as waterfall plots) introduce unique processing challenges. Each sensing node along the fiber measures the strain and outputs a corresponding time-series signal. By combining the time-series signals from multiple sensing nodes, a two-dimensional (2D) matrix is formed as a waterfall plot. As a result, one axis represents fiber distance (spatial dimension) and the other represents time (temporal dimension). This mechanism creates unique spatial-temporal coupling fundamentally distinct from natural images. Consequently, waterfall plots exhibit semantic abstraction analogous to acoustic streams rather than visually interpretable images. This characteristic renders conventional image processing techniques less effective for transforming raw waterfall plots into representations that reveal the underlying physical processes. Therefore, new deep-learning-based approaches are required to learn waterfall plots' abstract semantics or representations from a large amount of raw data.

Up to now, deep learning approaches for learning waterfall plots' representations have evolved through multiple paradigms. Supervised methods [13-14] leverage labeled datasets to learn latent representations, yet their effectiveness is constrained by limited labeled data. This limitation results in potential representation fragmentation. To address this constraint, semi-supervised methods emerged, integrating pseudo-labeling strategies [15] and consistency regularization [16] to use abundant unlabeled data. These techniques achieved superior performance in seismic phase picking [17] and signal recognition [18] compared to purely supervised approaches. However, their effectiveness diminishes when confronting distribution shifts between source domain (training dataset) and target domain (testing dataset). Recent efforts like the adaptive decentralized AI (ADAI) scheme [19] attempt to bridge this gap through domain adaptation techniques [20] by joint training on source-domain labeled data and target-domain unlabeled data. While ADAI alleviates distribution discrepancies, its representation learning remains heavily dependent on labeled data, introducing task-specific biases [21] and undermining generalization to novel signal types. On the other hand, self-supervised learning (SSL) methods have been utilized to eliminate label dependence by pre-training models entirely on unlabeled data. Ende et al. [22] proposed an approach for DAS by pre-training a U-Net [23], a convolution network, to reconstruct a blanked spatial column in noise-contaminated waterfall plots. Since the U-Net is primarily designed for image processing, its learned representations are suboptimal for DAS signals in the authors' opinion.

Recently, contemporary breakthroughs in large language [24-26] and vision models [27-28] highlight the transformative potential of Transformer architectures [29]. In this study, by adopting Transformers to SSL, we propose DAS-MAE for representation learning of 2D spatial-temporal waterfall plots. The core innovation of DAS-MAE lies in a spatial-temporal Transformer architecture combined with masking strategies. During pre-training, we implement *temporal-axis* masking that removes 50% sampling points in each waterfall plot. The proposed DAS-MAE encoder employs temporal self-attention modules to process visible unmasked sampling points across all spatial channels, generating compressed latent representations. In contrast, the DAS-MAE decoder adopts a lightweight Transformer structure that reconstructs the original signal from encoded representations. This small decoder design enforces the model's encoder to develop robust high-level sampling-to-semantics reasoning capabilities rather than relying on low-level contextual interpolation. After pre-training, DAS-MAE can cluster representations from the same type (e.g., walking, digging, etc.) of waterfall plots *without using any label information*. In few-shot learning experiments on an open dataset [30], our DAS-MAE model achieves a 1.1% error rate (64.5% relative improvement over the semi-supervised baseline) and demonstrates robust performance with minimal labeled data, reaching the critical 10% error threshold with only 90 labeled elements. Moreover, the pre-trained model is applied to a practical external damage prevention application, which consistently outperforms the from-scratch training model across three different amounts of training data. These results demonstrate both the effectiveness of pre-training and the model's high transferability among different applications and novel events.

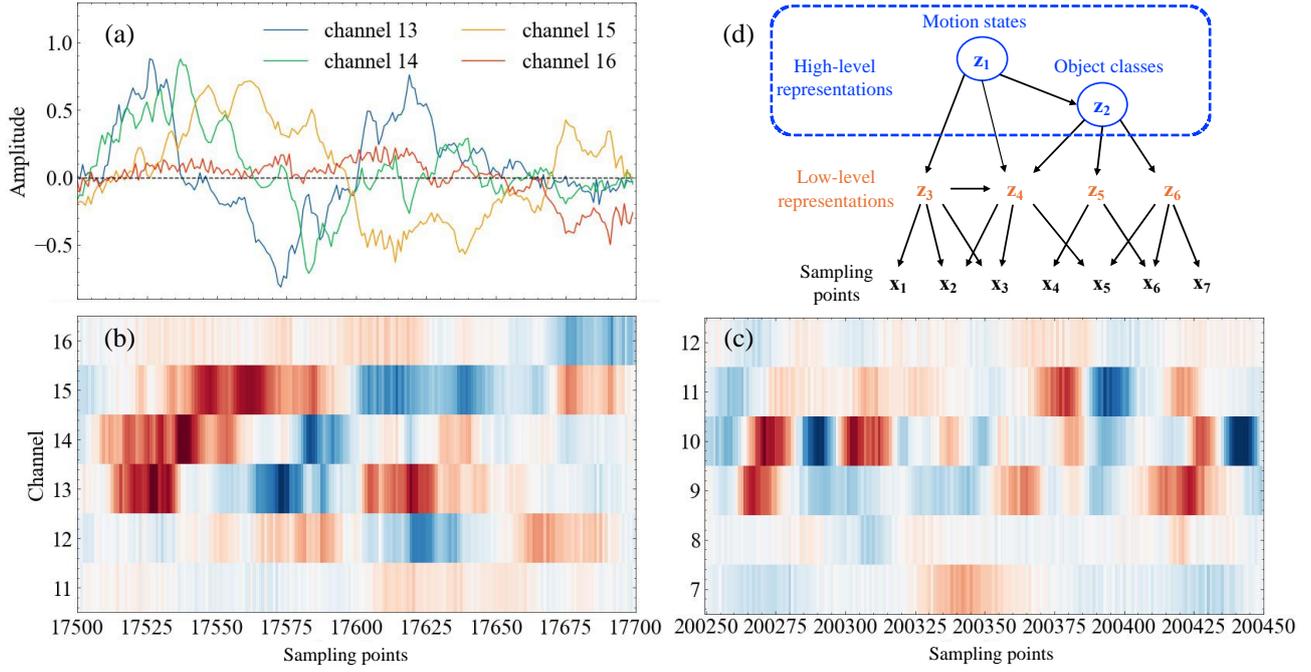

**Fig. 1 | Waterfall plot analysis and hierarchical data-generation structure.** (a) Spatial adjacent channels from a moving forklift waterfall plot. The waterfall plot depicts four adjacent spatial channels (13: blue, 14: green, 15: yellow, 16: red) recorded by a 2 kSa/s DAS equipment. While channels 13, 14, and 15 exhibit significant waveform similarity, this similarity disappears in channel 16. (b) 2D view of the forklift waterfall plot. This visualization includes the four channels shown in Fig. 1(a) and provides a comprehensive view of the spatial-temporal signal distribution across the monitored area. (c) Temporal evolution of the forklift waterfall plot. The waterfall plot after 90 seconds shows the center position shift from channel 14 to channel 10 (corresponding to 40 meters). (d) Hierarchical data-generation structure of waterfall plots. High-level representations generate both low-level representations and sampling points. While the forward process (high-level to low-level) is well-defined, the inverse inference (low-level to high-level) is non-trivial.

## Principles

### Waterfall plots

The DAS waterfall plot is formally represented as $\mathbf{x} \in \mathbb{R}^{C \times S}$, where $C$ denotes the number of spatial channels and $S$ represents temporal sampling points. The matrix point $x_{ij}$ encodes the instantaneous strain at the $i$-th spatial channel and $j$-th time sampling point. As shown in Fig. 1, localized disturbances (e.g., forklift movement) manifest as non-stationary spatial-temporal signals across adjacent channels and sampling points. The characteristics of the waterfall plot are listed as follows:

(a) **Asymmetric spatial-temporal information density**: A DAS waterfall plot exhibits temporal dominance

in information entropy, as evidenced by the superior performance of temporal-focused models: temporal 1D CNNs achieve 95.7% accuracy in 5-class event recognition [31], and 2D CNNs applied to time-frequency spectrum attain 88.2% accuracy for 10-class event classification [32]. In contrast, spatial-only models remain unexplored due to sparse spatial sampling (0.1 to 1 sampling points/meter versus $10^3$ to $10^4$ sampling points/second). Despite this asymmetry, spatial coherence plays an indispensable role in spatial-temporal dynamics. Although spatial information alone cannot differentiate event categories, it provides critical cues for inferring object motion states by tracing disturbance propagation across adjacent channels and sampling points. Ref. [33] confirms that integrating spatial information with temporal features enhances classification performance through spatial-temporal coherence modeling. Crucially, spatial information complements temporal features by encoding object kinematics rather than categorical distinctions. Thus, effective analysis demands cohesive spatial-temporal integration, where both dimensions are treated as mutually reinforcing components. Temporal features help event identification, while spatial features resolve motion trajectories. This interdependence underscores the necessity of unified frameworks that equitably leverage both domains for holistic DAS data analysis.

(b) **Short-term spatial-temporal coherence**: The DAS waterfall plot exhibits short-term spatial-temporal coherence, characterized by waveform similarity of time series at spatial-adjacent channels. As illustrated in Fig. 1(a), the time series $\{x_{ij} \,|\, 17500 \leq i \leq 17625, j = 14\}$ (green) displays near-identical amplitude information (a prominent peak spanning sampling points from 17500 to 17575 and a valley from 17575 to 17610) matching adjacent time series $j = 13$ (blue) and $j = 15$ (yellow) regardless of time delays. However, this similarity sharply degrades with the spatial distance increasing, as evidenced by the divergent waveform structure at $j = 16$ (red), where both peak and valley features vanish. This behavior aligns with findings in Ref. [34], which highlights that the "waveform coherence" (i.e., short-term spatial-temporal coherence) of the measured strain may not extend very far in space. The observed coherence validates the localized spatial-temporal correlation of disturbances, emphasizing the need for analysis frameworks that prioritize proximal channel-time relationships while accounting for rapid decorrelation at larger spatial offsets.

(c) **Long-term spatial persistence**: The DAS waterfall plot demonstrates long-term spatial persistence, where spatial information becomes critical for object tracking and motion state analysis over extended durations. As shown in Fig. 1(b) and (c), forklift-induced signals (spanning $\{x_{ij} | 12 \leq j \leq 15\}$ and $\{x_{ij} | 9 \leq j \leq 11\}$) persist across approximately 180000 temporal sampling points (90 seconds). Despite sharing similar signatures, distinguishing whether they originate from the same forklift remains challenging due to signal distortions caused by heterogeneous external factors. To resolve such ambiguities, long-term spatial-temporal persistence analysis is essential, which leverages spatial persistence across distant channels to disentangle intrinsic object signatures from transient disturbances. Specifically, spatial persistence enables the identification of consistent propagation trajectories (e.g., linear motion paths, directional trends) even when waveforms exhibit localized deviations. This capability underscores the necessity of integrating multiscale spatial features with temporal dynamics for robust object tracking.

(d) **Hierarchical data-generating structure**: Inspired by the hierarchical structure of images [35], we believe that waterfall plots exhibit analogous hierarchical structures. Fig. 1(d) illustrates the hierarchical data-generating structure defined by a directed acyclic graph (DAG) $\mathbf{G} = (\mathbf{V}, \mathbf{E})$, where $\mathbf{V}$ and $\mathbf{E}$ are the set of nodes and edges respectively. The node set $\mathbf{V} = (\mathbf{X}, \mathbf{Z})$ comprises waterfall plot's sampling points $\mathbf{X} = \{x_i\}_{i=1}^{C \times S}$ (flattened data) and latent representations $\mathbf{Z} = \{\mathbf{z}_j\}_{j=1}^{T}$ (Each $\mathbf{z}_j$ represents a multi-dimensional vector). Latent representation $\mathbf{z}_j$ and sampling point $x_i$ are generated by its parents as:

$$\begin{aligned} \mathbf{z}_j &= g_{\mathbf{z}_j}\left(\text{Pa}(\mathbf{z}_j)\right), \quad \forall j \in \{1,...,T\} \\ x_i &= g_{x_i}\left(\text{Pa}(x_i)\right), \quad \forall i \in \{1,...,C \times S\} \end{aligned} \tag{1}$$

where $g_{\mathbf{z}_j}$ and $g_{x_i}$ are invertible data generating functions, and $\text{Pa}(\cdot)$ denotes the parents of a certain node. In this theoretical framework, high-level representations (semantics) generate both low-level representations (elaborate and granular information) and sampling points. Conversely, inferring high-level semantics from low-level data constitutes a non-trivial inverse problem [36-37]. For waterfall plots, two distinct high-level representations emerge: object class $z_2$ (learned from waveform characteristics, e.g., identifying a "forklift" from localized vibrations) and motion state $z_1$ (inferred by identifying and tracking an object). Notably, while $z_1$ can causally inform $z_2$ as dynamic behavior often reflects object identity, the converse inference remains non-trivial. Lower-level representations, conversely, encapsulate signal attributes, including temporal fluctuations, short-term spatial-temporal coherence, and long-term spatial persistence.

**DAS-MAE**

Fig. 2(a) illustrates the DAS-MAE architecture for waterfall plot reconstruction through self-supervised masked modeling. The input waterfall plot $\mathbf{x} \in \mathbb{R}^{C \times S}$ is partitioned into non-overlapping rectangular patches $\{\mathbf{x}_i \in \mathbb{R}^{C_p \times S_p}\}_{i=1}^{N}$ where $N = \lfloor C/C_p \rfloor \times \lfloor S/S_p \rfloor$, and $\lfloor \cdot \rfloor$ is the floor function. Each patch $\mathbf{x}_i$ consists of $C_p$ channels ($C_p = 1 \leq C$) and $S_p$ temporal sampling points ($S_p = 624 \leq S$). During pre-training, we implement aggressive random masking by removing 50% of patches ($N_m = \lfloor 0.5N \rfloor$) following a uniform distribution (referred to as "random sampling"). The remaining $N_v = N - N_m$ visible patches $\mathbf{x}_{\mathbf{m}^c}$ (where $\mathbf{x}_{\mathbf{m}^c}$ denotes the complement of the mask set $\mathbf{m}$) are encoded by the DAS-MAE encoder into latent representations. A lightweight decoder estimates the masked patches $\hat{\mathbf{x}}_{\mathbf{m}}$ from representations and mask tokens. The reconstruction objective minimizes the normalized mean square error (MSE) across masked regions within the entire dataset $\mathcal{X}$:

$$L_{rec} = \mathbb{E}_{\mathbf{x} \sim \mathcal{X}} \left[ \frac{1}{N_m} \sum_{i=1}^{N_m} \left\| \mathbf{x}_{\mathbf{m}}^{(i)} - \hat{\mathbf{x}}_{\mathbf{m}}^{(i)} \right\|_2^2 \right]. \tag{2}$$

The 50% mask ratio (the ratio of removed patches) largely eliminates the waterfall plot's redundancy and prevents interpolation solutions. The lightwight decoder design forces the encoder to generate representations that invert the data generation hierarchy $x_i \rightarrow z_j$.

As illustrated in Fig. 2(b), the DAS-MAE framework employs a spatial-temporal Transformer (termed DAS-ViT) for both its encoder and decoder. The **encoder** processes visible patches through the following stages:

**Patch embedding**. The input patches $\mathbf{x}_{\mathbf{m}^c}$ are first flattened as $\mathbb{R}^{N_v \times C_p \times S_p \times D_i} \rightarrow \mathbb{R}^{N_v \times (C_p \cdot S_p) \times D_i}$, where $D_i = 1$ is an additional data dimension for network processing. The patch embedding layer then maps this to the

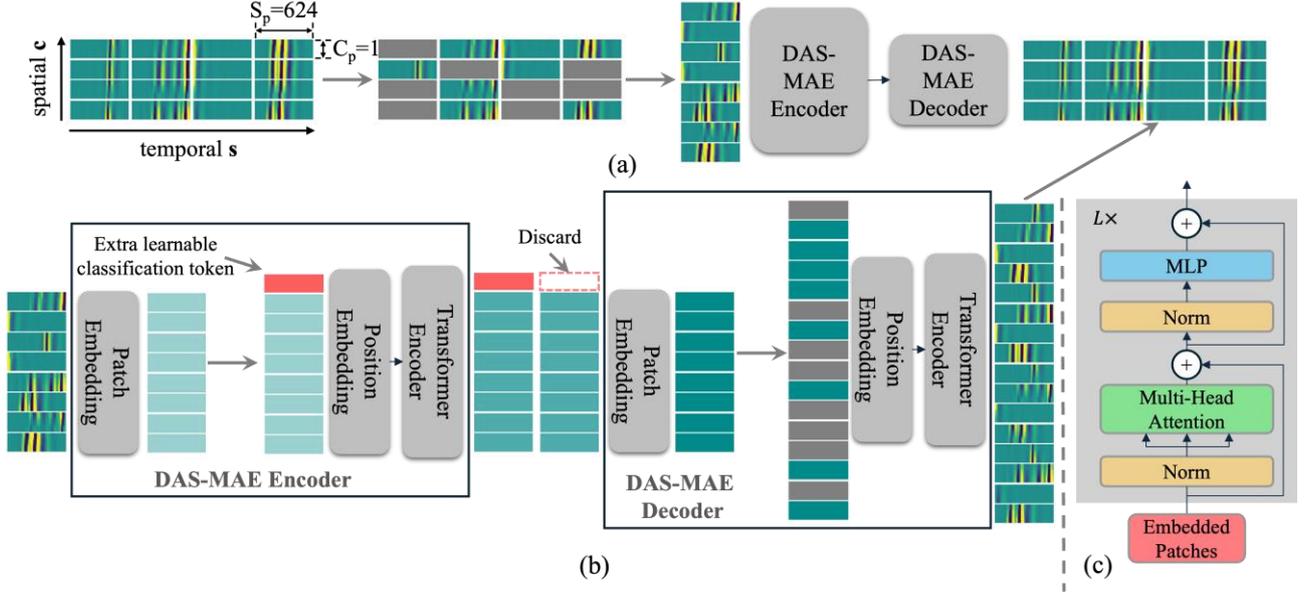

**Fig. 2 | Architecture of the DAS-MAE (DAS masked autoencoder) during pre-training. (a)** The overall structure of DAS-MAE. Its hyperparameters are given in Table 1. **(b)** The modules of DAS-MAE. It employs a DAS Vision Transformer (DAS-ViT) for both the encoder and decoder, each including a patch embedding layer, a position embedding layer, and a Transformer encoder. It is important to note that hyperparameter configurations of the encoder and the decoder differ significantly (see Table 1). The encoder is deliberately designed with a larger size compared to the decoder. **(c)** The structure of a Transformer encoder (depth of $L$, head of $H$) [38]. It consists of $L$ Transformer block, each including multi-head attention ($H$ heads), MLP block, and norm layers.

Transformer encoder dimension $D_e$, implemented by a convolution layer. **Position embedding**: A learnable classification token $\mathbf{x}_{\mathbf{CLS}} \in \mathbb{R}^{1 \times (C_p \cdot S_p) \times D_e}$ is prepended to the embedded patches. To retain positional information, standard 1D learnable position encodings [29] are added to the sequence. **Transformer processing**: The composite sequence (classification token and embedded patches) is processed by $L_e$ stacked Transformer blocks in Fig. 2(c). Each Transformer block contains a multi-head attention ($H_e$ heads), a MLP block, and norm layers. The encoder outputs high-level representations: $[\mathbf{z}_{\mathbf{CLS}}; \mathbf{z}_{\mathbf{m}^c}] \in \mathbb{R}^{(N_v+1) \times (C_p \cdot S_p) \times D_e}$. The DAS-MAE **decoder** operates exclusively on the encoded visible patches $\mathbf{z}_{\mathbf{m}^c}$ to reconstruct masked regions. Its workflow comprises: **Patch embedding**: A trainable linear layer first maps the dimension $D_e$ to the decoder's latent dimension $D_d$. **Position embedding**: The token sequence (length of $N_v$) is reconstituted into a complete set of $N$ tokens by replacing masked positions with a learnable and shared mask token (size of $\mathbb{R}^{1 \times (C_p \cdot S_p) \times D_d}$), indicating the presence of missing patches. Learnable 1D positional embeddings are injected to preserve spatial-temporal relationships. **Transformer processing**: The composite sequence undergoes $L_d$ Transformer blocks ($H_d$ heads). The decoder's output is reshaped and projected to match the original waterfall plot dimensions: $\mathbb{R}^{N\times(C_p\cdot S_p)\times D_d} \to \mathbb{R}^{N\times C_p\times S_p\times D_i}$. The tensors' shapes at each layer and values of model's hyperparameters are presented in Table 1.

Table 1 | Pretraining framework of DAS-MAE.

| Module/Variable | Sub-module | Layer | Hyperparameter | Output size/Size | Value |
|---|---|---|---|---|---|
| **x** (patched) | - | - | - | $\mathbb{R}^{N\times C_p\times S_p\times D_i}$ | $C_p = 1$ |
| $\mathbf{x_{m^c}}$ | - | - | - | $\mathbb{R}^{N_v\times C_p\times S_p\times D_i}$ | $S_p = 624$ |
| Encoder | Patch embedding | 1D Conv | $\mathbf{kernel} = C_p \times S_p, \mathbf{stride} = C_p \times S_p$ | $\mathbb{R}^{N_v\times(C_p\bullet S_p)\times D_e}$ | $N = 192$ |
| | Position embedding | - | - | $\mathbb{R}^{(N_v+1)\times(C_p\bullet S_p)\times D_e}$ | $N_v = 96$ |
| | Transformer encoder | Transformer block | $\mathbf{depth} = L_e, \mathbf{head} = H_e$ | $\mathbb{R}^{(N_v+1)\times(C_p\bullet S_p)\times D_e}$ | $D_i = 1$ |
| $[\mathbf{z_{CLS}; z_{m^c}}]$ | Encoder output | - | - | $\mathbb{R}^{(N_v+1)\times(C_p\bullet S_p)\times D_e}$ | $D_e = 576$ |
| $\mathbf{z_{m^c}}$ | Decoder input | - | - | $\mathbb{R}^{N_v\times(C_p\bullet S_p)\times D_e}$ | $D_d = 256$ |
| Decoder | Patch embedding | Linear | $\mathbf{in\_channel} = D_e, \mathbf{out\_channel} = D_d$ | $\mathbb{R}^{N_v\times(C_p\bullet S_p)\times D_d}$ | $L_e = 6$ |
| | Position embedding | - | - | $\mathbb{R}^{N\times(C_p\bullet S_p)\times D_d}$ | $H_e = 6$ |
| | Transformer encoder | Transformer block | $\mathbf{depth} = L_d, \mathbf{head} = H_d$ | $\mathbb{R}^{N\times(C_p\bullet S_p)\times D_d}$ | $L_d = 2$ |
| $\hat{\mathbf{x}}$ | - | - | - | $\mathbb{R}^{N\times C_p\times S_p\times D_i}$ | $H_d = 8$ |

The input patch design of DAS-MAE is tailored to align with the unique spatial-temporal characteristics of waterfall plots. As aforementioned, we observe an asymmetric spatial-temporal information density and significantly higher temporal redundancy than spatial redundancy. To prevent the model from learning trivial temporal interpolation in the reconstruction task, we utilize 1D temporal patches of shape $1\times 624$ (spanning 624 continuous temporal sampling points from a single spatial channel), instead of conventional square patches (e.g., $16\times 16$ in images). Masking such patches reduces temporal redundancy while preserving spatial dependencies, compelling the encoder to synthesize missing patches through both short-term spatial-temporal coherence and long-term spatial persistence. To operate this design, we adapt the Vision Transformer (ViT) architecture [38] for DAS and rename it as the DAS-ViT model. The key modification lies in the patch embedding layer: we employ 1D temporal convolutions with kernel and stride sizes of $1\times 624$ instead of using standard 2D convolutions. This ensures strict isolation of temporal patches across spatial channels, preventing spatial-temporal information leakage. The revised patch embedding mechanism enables the Transformer's self-attention layers to model inter-patch spatial-temporal relationships and learn high-level representations.

Effective self-supervised learning via masked autoencoder depends on calibrating reconstruction difficulty to induce an inverse hierarchical data-generation process for high-level representation learning. Specifically, the task must be sufficiently challenging to prevent the model from relying solely on interpolation (e.g., temporal redundancy exploitation or elaborate and granular information learning) yet not overly difficult to avoid underlearning fundamental signal characteristics. This balance ensures the encoder acquires necessary low-level representations in both number and quality for learning high-level representations or semantics (e.g., event categories). Moreover, the optimal reconstruction difficulty is signal-dependent: taking mask ratio (ratio of removed data) as an example, modalities with more inherent redundancy (e.g., images, 75% mask ratio [39]) tolerate aggressive masking, whereas less redundant signals (e.g., speech, 50% mask ratio [40]) require conservative strategies. DAS waterfall plots exhibit hybrid temporal redundancy and spatial sparsity, different from images and speech. We systematically ablate three critical factors for the optimal reconstruction difficulty:

**Mask ratio**: controls the proportion of patches removed. **Mask sampling strategy**: determines how masked patches are selected (random or others). **DAS-ViT architecture**, including patch shape (patch shape counteracts interpolation) and decoder capacity (lightweight decoders prevent excessive complexity while ensuring reconstruction fidelity). Our ablation studies (see ablation study section) quantify how these hyperparameters modulate task difficulty and influence representation quality.

**Evaluation protocol of pre-trained DAS-MAE**

To assess the semantic discriminability of DAS-MAE's learned representations (i.e., the representation quality), we evaluate DAS-MAE's performance on a classification task using labeled dataset $\mathbf{D} = \{(\mathbf{x}^{(k)}, \mathbf{y}^{(k)})\}_{k=1}^{K}$ where $\mathbf{y} \in \{0,1\}^M$ is a one-hot encoded label over $M$ event classes. Given an input $\mathbf{x} \in \mathbb{R}^{C \times S}$, the pre-trained encoder $\mathcal{A}$ generates latent representations $[\mathbf{z}_{\mathbf{CLS}}; \mathbf{z}] \in \mathbb{R}^{(N+1) \times (C_p \cdot S_p) \times D_e}$ *without masking*. Consequently, the representation length changes from $N_v$ to $N$ (detailed evaluation framework see Table S1, supplementary information). These representations are mapped to class probabilities $\hat{\mathbf{y}}$ via two distinct protocols.

Linear probing uses a lightweight linear classifier $\mathcal{D}$ (a single linear layer), which projects only the classification token's representation $\mathbf{z}_{\mathbf{CLS}}$ to class probabilities, while the encoder $\mathcal{A}$ remains frozen:

$$\arg\min_{\mathcal{D}} \mathcal{L}_{cls}\left[\mathbf{y}, \mathcal{D}\left([\mathcal{A}(\mathbf{x})]_0\right)\right], \quad (3)$$

where $\mathcal{L}_{cls}(\cdot)$ is the cross-entropy loss [41] for classification, and $[\mathcal{A}(\mathbf{x})]_0 = [\mathbf{z}_{\mathbf{CLS}};\mathbf{z}]_0 = \mathbf{z}_{\mathbf{CLS}}$. This method evaluates whether class-relevant representations emerge as linearly separable clusters in the latent space, which directly reflects the encoder's reconstruction-driven representation quality.

Fine-tuning attempts to use task-specific nonlinear features and optimizes both the encoder $\mathcal{A}$ and classifier $\mathcal{D}$ (a single linear layer). Here, all representations from patches $\mathbf{z} = [\mathcal{A}(\mathbf{x})]_{1:N}$ are aggregated via averaging before classification:

$$\arg\min_{\mathcal{A},\mathcal{D}} \mathcal{L}_{cls}\left[\mathbf{y}, \mathcal{D}\left(\underset{(1\text{st dim})}{\text{Mean}}\left([\mathcal{A}(\mathbf{x})]_{1:N}\right)\right)\right]. \quad (4)$$

This method demonstrates the model's adaptability. It leverages minimal task-specific training (far fewer epochs than supervised training from scratch) to maximize transfer performance.

# Experiments

DAS-MAE ($\sim$ 28M parameters) was pre-trained on an open waterfall plot dataset, contributed by X. Cao, et al. [30], comprising $\sim$ 15000 elements (12 channels $\times$ 10000 temporal sampling points per data) acquired at sampling rates of 12.5 kHz or 8.0 kHz. The dataset encompasses six distinct event classes: background noise, digging, knocking, watering, shaking, and walking. The training set contains about 12000 elements ($\sim$ 2000 elements for each event), while the testing set contains 3000 elements ($\sim$ 500 elements for each event). Pre-training DAS-MAE leveraged only raw data without class labels. We adopted a batch size of 64 and trained for 500 epochs using the AdamW optimizer [42] with cosine learning rate decay [43] (initial learning rate of $10^{-3}$, 40-epoch warmup). The entire pre-training was completed in about 6 hours on an NVIDIA RTX 4090 GPU. The pre-trained model is able to achieve real-time inference (2 ms @ RTX 4090), which is suitable for field deployment.

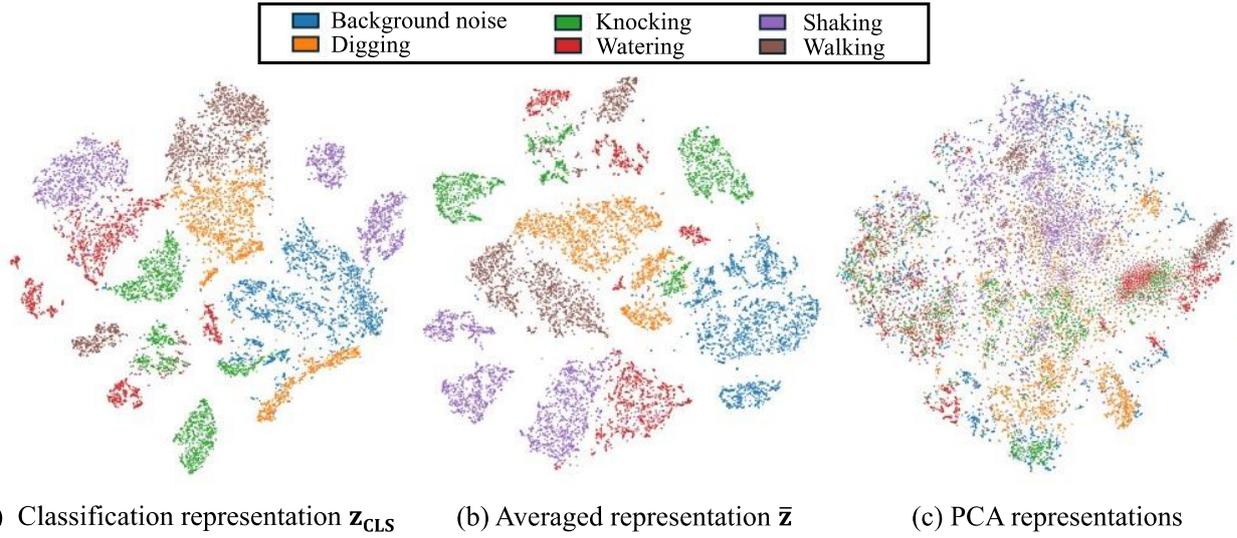

(a) Classification representation $\mathbf{z_{CLS}}$   (b) Averaged representation $\bar{\mathbf{z}}$   (c) PCA representations

**Fig. 3 | The t-SNE visualization of representations learned by DAS-MAE and principal component analysis (PCA) on open dataset** [30]. Labels are used for color assignment to evaluate unsupervised learning of high-level representations. **(a)** Classification representation $\mathbf{z_{CLS}}$ from DAS-MAE. **(b)** Averaged representations $\bar{\mathbf{z}}$ from DAS-MAE. **(c)** Representations learned from PCA.

**Visualization of learned representations**

We began with an intuitive and qualitative assessment of DAS-MAE's learned representations, visualizing them via the t-distributed stochastic neighbor embedding (t-SNE) [44]. The t-SNE is a nonlinear dimension reduction technique that preserves the clustering information in the high-dimensional space and reveals it in the low-dimensional visualization. Using the pre-trained encoder, we extract two representation sets from the training dataset [30]: the classification representations $\mathbf{z_{CLS}}$ and averaged representations $\bar{\mathbf{z}} = \underset{(1\text{st dim})}{\text{Mean}}(\mathbf{z})$. Both $\mathbf{z_{CLS}}$ and $\bar{\mathbf{z}}$ are projected to 2D using t-SNE (perplexity=40, learning rate=2000, 500 iterations) and visualized with class-colored scatterplots. Note that the class (label) information only assigned colors in Fig. 3 for better visual distinction.

As shown in Fig. 3(a), the classification representations $\mathbf{z_{CLS}}$ exhibit well-separated clusters for all six event categories, with minimal overlap between distinct classes. Notably, the background noise class (blue) forms a tight cluster, reflecting its stationary characteristics. For dynamic events (walking, knocking, etc.), intra-class dispersion arises primarily from unannotated subcategories. For instance, the 'walking' class (brown) encompasses running and walking events, which DAS-MAE disentangles into subclusters. This problem can be effectively solved by fine-tuning with given labels, as evidenced by the less 1% error rate in ablation study section. Comparatively, averaged representations $\bar{z}$ show similar separation in Fig. 3(b). To validate the superiority of DAS-MAE's representations, we contrast them with representations from principal component analysis (PCA) [45] (implemented by a linear autoencoder [46]). As shown in Fig. 3(c), the PCA method fails to distinguish different event types since the generated representations all overlap. These visualizations empirically validate that DAS-MAE learns semantically meaningful representations without label supervision.

Table 2 | Error rate of different models.

| Labeled data No. | ACAB [18] | DAS-MAE | Relative improvement |
|---|---|---|---|
| 90 | 16.5% | **10.0%** | 39.3% |
| 240 | 6.2% | **3.6%** | 41.9% |
| 480 | 4.4% | **2.8%** | 36.4% |
| 960 | 3.8% | **1.3%** | 65.8% |
| 1230 | 3.1% | **1.1%** | 64.5% |

**Quality of learned representations**

To quantitatively evaluate the quality of learned high-level representations, we compared our DAS-MAE model with the semi-supervised ACAB (ACNN-SA-BiLSTM) model [18] on the open dataset [30]. ACAB was trained using the Mean Teacher framework [16] with limited labeled data subsets (masking labels for most training data [47]), where DAS-MAE was fine-tuned under identical labeled data constraints (i.e., few-shot learning). DAS-MAE employed AdamW optimization [42] with a cosine learning schedule (initial learning rate=$10^{-5}$, weight decay=0.05, 4 warm-up epochs) over 50 fine-tuning epochs. As shown in Table 2, DAS-MAE demonstrates superior classification performance across all limited labeled-data numbers. With only 90 labeled elements (15 per class), DAS-MAE achieves a 10.0% error rate (ER), outperforming ACAB (16.5% ER) by 39.3% relative improvement (RI), calculated as:

$$\mathrm{RI} = \frac{\mathrm{ER}_{\mathrm{ACAB}} - \mathrm{ER}_{\mathrm{DAS\text{-}MAE}}}{\mathrm{ER}_{\mathrm{ACAB}}} \times 100\%. \tag{5}$$

Notably, DAS-MAE reaches the 10% ER threshold critical for practical DAS deployment with minimal labels, whereas ACAB remains unsuitable under such constraints. With 1230 labeled data, DAS-MAE attains a record 1.1% ER versus ACAB's 3.1%, yielding 64.5% RI. These results highlight that both models benefit from additional labeled data, but DAS-MAE's advantage scales more prominently (the increment of RI in Table 2), suggesting a superior representation learning ability from pre-training.

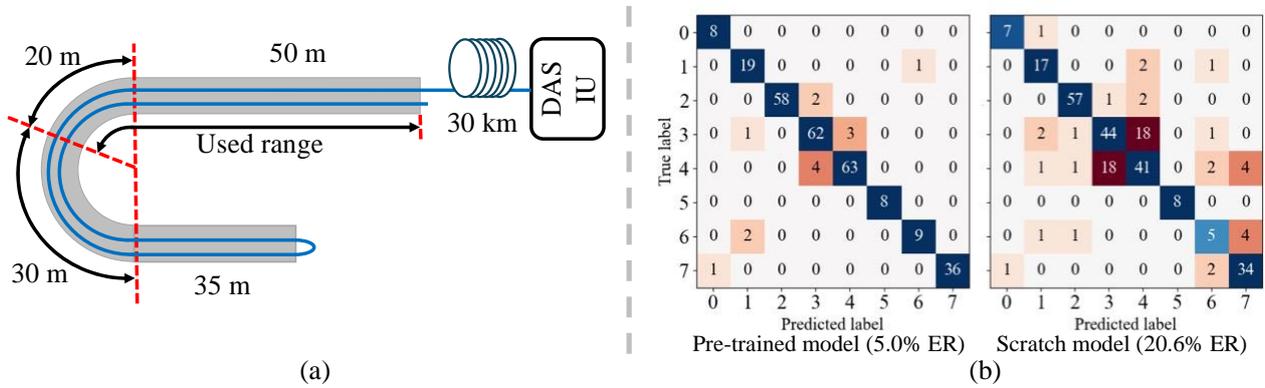

**Fig. 4 | Generalization of DAS-MAE learned representations in the external damage prevention experiment. (a)** Experimental layout. The gray part is a trench, where the sensing fiber (blue line) is laid out in a back-and-forth configure. There is a 30-kilometer fiber between the trench and the DAS interrogator unit (DAS IU). The vibration data are sampled in the 'used range' of fiber (i.e., the first 70 meters). **(b)** Confusion matrices of DAS-MAE by pre-training (pre-trained model) and from-scratch training (scratch model) with all (100%) training data. The matrices are structured as: The horizontal axis represents the predicted labels, while the vertical axis denotes the true labels. Each cell $(i, j)$ in the matrix contains the number of elements from class $i$ that were classified as class $j$.

### Generalization of learned representations

To further evaluate the generalization of learned representations in practical scenarios, we conducted field experiments in an external damage prevention application deployed in Zhengzhou, China. The experimental setup (Fig. 4) employed a U-shaped fiber trench with three geometrically distinct sections, monitored by a DFVS-850 DAS system operating at 2 kSa/s with 10-meter spatial resolution (detailed information see supplementary information). The compiled dataset contains eight vibration event classes: background noise (0), road roller in motion (1), road roller in compaction (2), excavator in excavation (3), excavator in motion (4), electric drill in operation (5), fully-loaded forklift in motion (6), and unloaded forklift in motion (7), which are distinct from the vibration types using in model pre-training. Our dataset contains ～1400 elements, and the number of elements in each type is 42 (0), 100 (1), 300 (2), 332 (3), 334 (4), 38 (5), 54 (6), 186 (7), respectively. This dataset exhibits significant class imbalance, with excavator in motion/excavation events occurring at an order-of-magnitude higher frequency than background noise, electric drill operations, or fully-loaded forklift activities. Each type of waterfall plot is divided into training and testing data in a 4:1 ratio, creating the training and testing sets.

Table 3 | Error rate of DAS-MAE in practical external damage prevention application.

| Training data (portion \| volume) | Pre-trained model | Scratch model (w/o pre-training) | Relative improvement |
| --- | --- | --- | --- |
| 100% \| 1100 | **5.0%** | 20.6% | 75.7% |
| 62.5% \| 700 | **12.4%** | 23.5% | 47.2% |
| 25% \| 280 | **16.2%** | 26.6% | 39.1% |

The pre-trained DAS-MAE was fine-tuned using three portions of the training set: 100% (∼ 1100 elements), 62.5% (∼ 700 elements), and 25% (∼ 280 elements). Using all training data (100%), the pre-trained model achieved a 5.0% test error rate, exceeding the practical deployment threshold by 5%. This result also demonstrates 15.6% absolute improvement and 75.7% relative improvement over from-scratch training (termed as the scratch model). Fig. 4(b) shows the confusion matrices of the pre-trained model and scratch model respectively. The pre-trained model makes much fewer prediction errors in full-load forklift operation (6) than the scratch model, which indicates that pre-training effectively mitigated class imbalance effects. Additionally, the confusion between excavator in excavation (3) and excavator in motion (4) in the scratch model is clarified in the pre-trained model. Under limited training data conditions (62.5% and 25% training set), the pre-trained model maintained 12.4% and 16.2% error rates respectively, consistently outperforming the scratch model by ∼ 10% absolute improvement and over 39% relative improvement (Table 3). These results demonstrate the effectiveness of the DAS-MAE pre-training phase and the substantial performance enhancement it yields. This further proves the high transferability of our model in different applications with novel event types.

## Ablation study

We conducted systematic hyperparameter ablation studies on DAS-MAE to optimize its configuration for waterfall plot characteristics and evaluate its representation learning capacity. The ablation protocol follows a two-stage process: First, we pre-trained DAS-MAE models with individual hyperparameters using the same training schedules (epochs, learning rates, etc.) in experiment section. Subsequently, we evaluated the learned representations through fine-tuning and linear probing approaches on the open classification dataset [30], using classification accuracy on its testing set as the quantitative metric for representation quality. To ensure controlled comparisons, each ablation experiment modified only one hyperparameter while preserving all other architectural components.

Table 4 | Ablation studies on DAS-MAE hyperparameter (measured by error rate).

(a) Ablation of patch shape.

| Patch shape | Recon. loss on testing set | Linear probing | Fine-tuning |
|---|---|---|---|
| $1 \times 624$ | 0.0023 | **2.45%** | **0.32%** |
| $2 \times 312$ | 0.0025 | 7.02% | 1.88% |
| $4 \times 156$ | 0.0021 | 4.38% | 1.05% |
| $6 \times 104$ | 0.0019 | 6.51% | 0.64% |
| $12 \times 52$ | 0.0023 | 2.74% | 0.61% |

(b) Ablation of mask ratio.

| Mask ratio | Linear probing | Fine-tuning |
|---|---|---|
| 30% | 4.87% | 1.31% |
| 40% | 4.57% | 0.57% |
| 50% | **2.45%** | **0.32%** |
| 55% | **2.45%** | 0.61% |
| 60% | 2.61% | 0.86% |
| 70% | 2.55% | 0.92% |
| 80% | 3.06% | 1.34% |

(c) Ablation of mask strategy.

| Mask strategy | Linear probing | Fine-tuning |
|---|---|---|
| Random sampling | **2.45%** | **0.32%** |
| Temporal sampling | 8.73% | 3.18% |
| Spatial sampling | 5.41% | 1.50% |

(d) Ablation of decoder shape.

| Decoder depth | Fine-tuning | Decoder width | Fine-tuning |
|---|---|---|---|
| 1 | 0.54% | 128 | 0.70% |
| 2 | **0.32%** | 256 | **0.32%** |
| 3 | 1.56% | 384 | 0.41% |

**Patch shape**

To investigate the unique spatial-temporal characteristics in DAS waterfall plots, we conducted controlled experiments maintaining a constant volume of patch sampling point $C_p \times S_p = 624$ while varying the patch shape of DAS-ViT. As quantified in Table 4a, the $1 \times 624$ temporal-patch configuration achieved optimal performance (2.45% linear probing error rate and 0.32% fine-tuning error rate) by effectively leveraging short-term spatial-temporal coherence. The significant 4.5% linear probing performance degradation is observed in $2 \times 312$ multi-channel patches. It stems from disrupted short-term spatial-temporal coherence since the waveform similarity beyond two adjacent channels sharply degrades (the analysis in experiment section: waterfall plot). The absence of short-term spatial-temporal coherence makes the reconstruction task so challenging that the pre-trained model learns insufficient low-level representations in the hierarchical data-generating structure and fails in high-level representation learning. Progressive temporal context reduction (from $2 \times 312$ to $12 \times 52$) partially mitigated this through increased inter-patch temporal correlation (i.e., an easier reconstruction task with higher inter-patch temporal redundancy). The $12 \times 52$ configuration attains suboptimal performance (2.74% linear probing error rate and 0.61% fine-tuning error rate). However, the performance gap to the $1\times 624$ patch shape underscores the importance of short-term spatial-temporal coherence, which can not be compensated by using an easier reconstruction task.

Moreover, the proposed $1 \times 624$ patch design adheres to the information redundancy principle for MAE-based architecture across data modalities. This principle dictates that patch shape should correlate with local information redundancy along each dimension. In image processing [39] where information redundancy is isotropic, square $16\times 16$ patches optimally balance local context. Video signals compromise an additional temporal dimension with lower information redundancy than both spatial dimensions, where the square patches are extended to $16\times 16\times 2$ patches [48]. DAS waterfall plots exhibit extreme spatial-temporal redundancy asymmetry, with temporal redundancy significantly surpassing spatial redundancy. This characteristic distinguishes them from video sequences and aligns them more closely with microphone array recordings. Consequently, we use the patch size of $1\times 624$ to accommodate more redundant temporal information and balance the spatial and temporal redundancy.

**Mask ratio**

The mask ratio constitutes a critical hyperparameter in DAS-MAE that regulates the difficulty of the reconstruction task and determines the abstraction level of learned representations. Both overly conservative (30%) and aggressive (80%) masking strategies result in suboptimal low-level representations within the hierarchical data structure, as evidenced by their significantly elevated error rates in Table 4b. The ablation study reveals an optimal intermediate mask ratio of 50%. It achieves minimal error rates of 2.45% (linear probing) and 0.32% (fine-tuning), suggesting that masking half patches provides the most appropriate reconstruction task difficulty. Notably, the model exhibits differential sensitivity to mask ratios across evaluating approaches. The fine-tuning approach demonstrates robust performance across a broad mask ratio range (40%-70%), while linear probing shows a substantial 2% error rate reduction between 40% and 50% masking. This discrepancy highlights the practical advantage of fine-tuning for enhanced stability in practical.

The observed 50% optimal mask ratio for waterfall plots aligns with speech signal processing benchmarks (50% mask ratio) while contrasting with the higher 75% mask ratio typically used in image processing. This comparative analysis implies that waterfall plots exhibit intermediate redundancy characteristics: their temporal redundancy resembles 1D speech signals, yet their spatial redundancy remains significantly lower than conventional 2D images. The similar mask ratio between 1D temporal speech signal and 1D temporal × 1D spatial waterfall plots indicates the dominance of temporal information in DAS signals. Such findings position waterfall plots as a distinct modality in the redundancy aspect between 1D audio and 2D visual data.

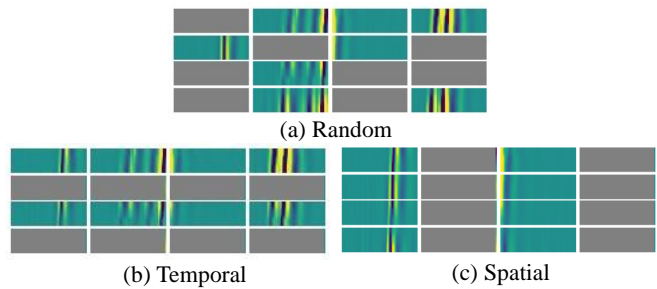

(a) Random

(b) Temporal       (c) Spatial

**Fig. 5 | Mask sampling strategies determine the difficulty of the pre-training reconstruction task, influencing the reconstruction quality and representations. (a)** Random sampling (our default) masks the patches under a uniform distribution. **(b)** Temporal sampling removes the entire row of patches. **(c)** Spatial sampling removes the entire column of patches.

### Mask sampling strategy

We systematically evaluated three distinct mask sampling strategies (Fig. 5) to elucidate their impact on DAS-MAE's representation learning capabilities. Temporal sampling in Fig. 5(b), which removes entire spatial-channel patches, forces reconstruction through spatial-adjacent waveform similarity alone (no useful temporal redundancy). Despite this constraint, DAS-MAE achieved competitive error rates of 8.73% (linear probing) and 3.18% (fine-tuning). This underscores the significance of spatial correlations in waterfall plots such as short-term spatial-temporal coherence and long-term spatial persistence. In contrast, spatial sampling (Fig. 5(c)) eliminated temporal redundancy by masking column-wise patches, yielding better performance with 5.41% linear probing error rate (38% RI compared to temporal sampling) and 1.50% fine-tuning error rate (53% RI). This marked improvement confirms the temporal dominance in DAS information entropy, where temporal information provides more high-level representations. Random sampling (Fig. 5(a)) proved optimal, achieving 2.45% linear probing error (55% RI over spatial sampling) and 0.32% fine-tuning error (79% RI), as quantified in Table 4c. The random masking strategy enables simultaneous exploitation of both spatial-temporal correlations while introducing beneficial randomness that enhances model robustness against overfitting.

**Decoder design**

Ablation studies are conducted to delineate the decoder's role in DAS-MAE representation learning, which reconstructs the original waterfall plots from the learned representations in the reconstruction task. We believe that the decoder fundamentally mediates the encoder's representation learning trajectory through its reconstruction capability. In other words, when the decoder's parameter count is reduced (weakening its reconstruction ability), the encoder tends to generate representations that make the reconstruction task easier for the decoder (i.e., the learned representations are more optimized for the reconstruction task). Conversely, the encoder tends to generate representations that are not easy to reconstruct when the decoder is large and capable. These learned representations are more optimized for the classification task. On the other hand, an overly large decoder would also reduce the encoder's representation learning ability. As a result, linear probing is too simplistic to demonstrate the quality of learned representations (whether it is high-level representation) due to its frozen encoder constraints. While fine-tuning the entire encoder offers greater robustness to this issue and more clearly reveals the transferability of representations across diverse tasks. Consequently, only the fine-tuning results offer meaningful insights during decoder ablation.

Table 4d varies the depth $L_d$ and width $D_d$ of the decoder separately. The width was varied while maintaining a fixed depth $L_d = 2$. Conversely, when manipulating the depth, the width remained constant at $D_d = 256$. The optimal performance of 0.32% error rate is achieved with a depth of 2 and a width of 256 ($\{L_d = 2, D_d = 256\}$). It is worth noting that employing a larger decoder (e.g., $\{L_d = 4, D_d = 256\}$ or $\{L_d = 2, D_d = 384\}$) may lead to a less optimal representation. Since the enhanced reconstruction capabilities of a more powerful decoder exist, the encoder does not attempt to learn the most effective representation in the reconstruction task. Surprisingly, even with an extremely small decoder $\{L_d = 1, D_d = 256\}$, the model achieves a remarkable performance of 0.54% error rate. It's important to note that a single Transformer block is the minimum requirement for propagating information from learned representations to estimated waterfall plots. This result further indicates that a large decoder is not essential for learning high-level representations of waterfall plots.

## Discussion

This section outlines the comparative advantages of distributed fiber-optic sensing over conventional single-point measurement methodologies. Conceptually, DAS constitutes a dense array of spatially coordinated vibration sensors, where each temporal sequence in waterfall plots corresponds to data recorded by an individual single-point sensor. Leveraging the open dataset [30], we construct two specialized datasets: a pre-training dataset $\mathbf{D}_{pre}$ containing every temporal sequence in waterfall plots, and a classification set $\mathbf{D}_{cls}$ comprising maximum-energy sequences per waterfall plot. We implement an architecturally identical model $\mathcal{M}$ to ensure parity with DAS-MAE training conditions. The model $\mathcal{M}$ was pre-trained on $\mathbf{D}_{pre}$ and fine-tuned on $\mathbf{D}_{cls}$, maintaining consistent hyperparameters and training settings to DAS-MAE throughout. The pre-training difference between model $\mathcal{M}$ and DAS-MAE lies in the input data shape, where $\mathcal{M}$ accepts temporal sequences (i.e., $C = 1$) to simulate the signals from a single-point sensor (detailed information see supplementary information).

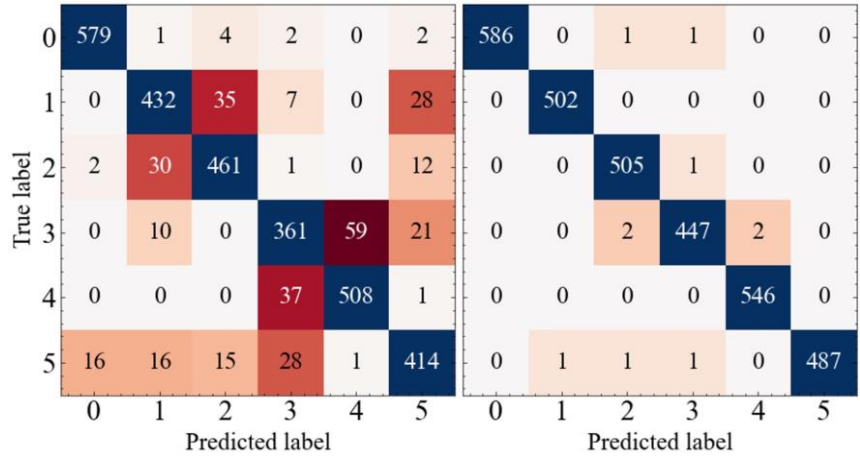

**Fig. 5 | Confusion matrices of the single-point vibration sensor and DAS from the DAS-MAE framework.** The matrix is structured as in Fig. 4(b). The denotes correspond to background noise (0), digging (1), knocking (2), watering (3), shaking (4), and walking (5). Left: single-point vibration sensor with 11.23% error rate. Right: distributed acoustic sensor with 0.32% error rate.

The empirical comparison reveals a substantial performance disparity: DAS-MAE (distributed sensing) achieves 0.32% classification error compared to an 11.23% error rate for model $\mathcal{M}$ (single-point sensing), as quantified in Fig. 6 through confusion matrix analysis. This 10.91% error rate decrement or 97.07% relative improvement demonstrates the advances of distributed sensing over single-point sensing. The spatial-temporal integration inherent to DAS architecture addresses three critical limitations of single-point detection: First, distributed configurations enable cross-validation between adjacent sensors, mitigating individual node misclassifications through majority consensus. Second, the simultaneous acquisition of temporal sequences across multiple spatial channels provides complementary signal features unavailable in isolated measurements. Third, multi-node correlation analysis reveals latent physical interactions between vibration sources and propagation media and provides long-term spatial persistence, enhancing representation discriminability. The experimental results validate that spatial correlation exploitation fundamentally enhances detection robustness beyond temporal analysis limitations inherent in single-point sensor systems.

## Conclusions

In conclusion, this work presents the DAS-MAE, which is the first self-supervised Transformer-based framework adopting masked autoencoders for distributed acoustic sensing, to the best of our knowledge. By reconstructing masked waterfall plots, the pre-trained model learns transferable and high-performance representations that encode the 'language' of DAS-recorded vibrations. This language adheres to a hierarchical data-generating structure, ranging from localized events or the propagation of time-series arrays to latent representations. For tasks requiring a higher understanding level of waterfall plots, our model can interpret task-specific data and provide comprehensive representations for enhanced performance. As a result, DAS-MAE contributes to the development of foundation models for intelligent distributed acoustic sensing. Additionally, the pre-trained model generates representations based on learned statistics from massive unlabeled waterfall plots. These representations reveal inherent characteristics and biases in the data, allowing human operators to gain deeper insights into the hierarchical data-generating structure and abstract waterfall plots. Such insights enhance the potential for broader application of DAS equipment in various fields. In turn, improved human insight enhances the interpretability of DAS-MAE, establishing a positive feedback loop that ultimately improves the transparency of our deep learning model.

# Acknowledgements

This work was supported by National Natural Science Foundation of China (NSFC) under Grant U2239203 and 62005163.

# Author contributions

Junyi Duan proposed the original idea, did formal analysis and wrote original draft. Jiageng Chen supported the methodology, validated the method and reviewed & edited the draft. Zuyuan He proposed the idea, supervised the project and reviewed & edited the draft.

# Competing interests

The authors declare no competing financial interests.